\documentclass[english]{article}
\usepackage[T1]{fontenc}
\usepackage[latin9]{inputenc}
\usepackage{geometry}
\usepackage{float}
\usepackage{booktabs}
\usepackage{textcomp}
\usepackage{color}
\usepackage{graphicx}
\usepackage{setspace}
\usepackage{esint}

\usepackage{babel}

\makeatletter
\def\@captype{figure}
\makeatother

\begin{document}

\title{A Demonstrator for Bolometric Interferometry}

\author{\small Adnan GHRIBI$^*$, Andrea Tartari$^{*\dag}$, Silvia Galli$^*$, Michel Piat$^*$, Eric Breelle$^*$, Jean-Christophe Hamilton\thanks{AstroParticule \& Cosmologie, Université Paris Diderot} ,\\
\small Sebastiano Spinelli$^\dag$, Massimo Gervasi$^\dag$, Mario Zannoni\thanks{Radio Group, Università degli Studi di Milano - Bicocca} .}

\maketitle

\begin{abstract}
Bolometric Interferometry (BI) is one of the most promising techniques
for precise measurements of the Cosmic Microwave Background polarization.
In this paper, we present the results of DIBO (Démonstrateur d'Interférometrie
Bolométrique), a single-baseline demonstrator operating at 90 GHz,
built to proof the validity of the BI concept applied to a millimeter-wave
interferometer. This instrument has been characterized in the
laboratory with a detector at room temperature
and with a 4 K bolometer. This allowed us to measure
interference patterns in a clean way, both (1) rotating the source
and (2) varying with time the phase shift among the two interferometer's
arms. Detailed modelisation has also been performed and validated with measurements.
\end{abstract}
%\newpage{}

\section*{Introduction}

\subsection*{Scientific Background}

The Cosmic Microwave Background (CMB) is a 3K black body radiation
last scattered about $300000$ years after the Big Bang. Dedicated instruments
have observed CMB anisotropies both in temperature (of the order of
$10^{-5}$) and in polarization. These fluctuations contain precious
cosmological information concerning the contents and the dynamics
of our Universe. Nevertheless, most of the polarized information remains
to be discovered. This is especially the case for the so called B-modes
of CMB polarization that are expected to be at least 3 orders of magnitude
lower than temperature anisotropies \cite{key-1}. B-modes could contain the finger
print of the hypothetical primordial gravitational waves predicted
by Einstein's General Relativity. This is is a major challenge in
cosmology in which several research groups are involved.

\subsection*{Instrumentation for cosmology \& Bolometric Interferometry}

Bolometers, cooled to very low temperatures (<0.3K) are the most sensitive
detectors for large bandwidth detection in the sub-millimeter and
millimeter wavelength range. Their principle of operation is to convert
the power carried by the electromagnetic waves into heat through an
absorber and to measure its temperature. The measured signal is directly
proportional to the incoming power. Many experiences dedicated to
CMB measurement are based on bolometric detection (Cobe-FIRAS, Boomerang,
Maxima, Archeops, Planck HFI, among others\footnote{see for example \cite{key-2} for a comprehensive list of experiments}). The future ESA Planck mission \cite{key-3} will use spiderweb bolometers which will operate near the photon noise limit. However, the detection of the B modes requires an improvement of up to two orders of magnitude in sensitivity with respect to the Planck mission. The only solution is therefore to increase the number
of detectors.

The CMB polarization characterization is nevertheless not only a question
of sensitivity. It also needs an exquisite control of all the instrumental
and systematic effects. As an example, for an imager, any lenses or
mirrors used to define the beam in the sky are sources of spurious
polarization effects that can mask the tiny B-mode signal.

Using an interferometer is an interesting alternative to imager \cite{key-4}. It
directly measures the Fourier transform of the intensity and polarization
distribution on the sky. Since only correlated signals are detected,
such system is intrinsically less sensitive to parasitic effects.
Many groups choosed this direction for the measurement of the CMB
polarization. Some interferometers used for that purpose are DASI,
CBI and VSA \cite{key-5}. These experiments use coherent detection techniques where
the signal received is amplified before being detected. The amplification
is intrinsically noisy leading to less sensitive instruments. Moreover,the correlation need a narrow bandwidth that further reduce the sensitivity.

In order to reach the bolometric sensitivity together with the systematic
cleanness of interferometers, we are developing an instrumental concept
based on bolometric interferometry : we use adding interferometry
combined with bolometric detection. This instrumental concept will
be applied to a project called BRAIN/MBI (B Mode Radiation Interferometer / Millimeter-wave Bolometric Interferometer) \cite{key-6}.
This project is an international collaboration targeting the detection
and characterization of the B modes with a ground based experiment
in the Concordia base (Antarctica). The final instrument will work
on three frequency bands (90 GHz, 150 GHz, 220 GHz) with 30\% bandwidth
for each band. This is needed to subtract the foregrounds and recover a maximum power from the
spectrum. The full instrument
will be made of 9 modules, each module being
made of $12\times 12$ horns. A set of phase shifters in a well defined
architecture will allow us to retrieve all the correlations between
the different baselines. A first step in the development of such instrument
is the proof of concept. For that reason, a simplified instrument
DIBO (Démonstrateur d'interférométrie bolométrique) was built with commercial quasi-optical components.

In the first section of this paper, we begin by introducing the instrument and its working principle. We then show its properties and some useful tools for the analysis of the results. The third section describes the different experimental setups developed for DIBO. In the fourth section, the results are shown both for bolometric and power-meter measurements. We finally conclude on perspectives for a next generation bolometric interferometer.

%\newpage{}

\section{The instrument}

\begin{center}
%\begin{figure}
\begin{centering}
\includegraphics[bb=137bp 380bp 339bp 740bp,clip,width=4cm]{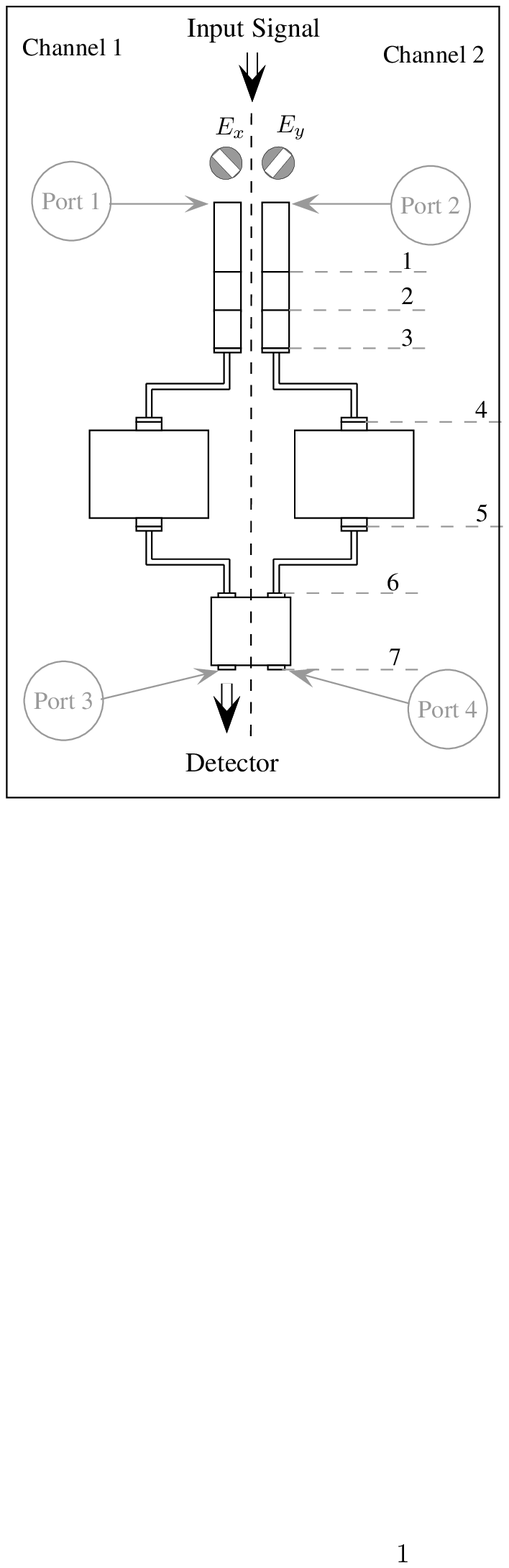}
\par\end{centering}
\begin{centering}
\caption{General layout of the instrument design : (1) two corrugated horns, (2) two circular to rectangular transitions, (3) two +/-45\textdegree{} twists, (4 and 6) four bends, (5) two controllable phase shifters, (7) 90\textdegree{} hybrid coupler.}
\label{fig:general layout}
\par\end{centering}
%\end{figure}
\par\end{center}

\subsection{Instrument description}

The DIBO instrument, as described in figure \ref{fig:general layout}, has two horns pointing toward the same direction. Afterwards, the incoming electromagnetic wave pass through a circular to rectangular
(WR10) waveguide transition. The output of this transition allows
the selection of a single polarization direction. The two E-planes
outputs are orthogonal. After a $\pm45$\textdegree{} twist, the waves are phase shifted with controllable phase shifters before being
combined by means of a 90\textdegree{} hybrid coupler. The signal
is then detected in one at the outputs of the coupler.

\subsection{Working principle}

The horns are assumed to have the same beam $B(\vec{s})$. They receive
the same signals, except that one is out of phase by $\phi_{path}=2\pi\vec{D\cdot}\vec{s}/\lambda$
compared to the other, where $\vec{D}$ is the separation between
the two horns (the baseline), $\vec{s}$ the unitary direction
of the incoming radiation and $\lambda$ the wavelength. This phase
shift is purely geometrical and is due to the different optical path
to reach each horn antennas. At the output of the antennas, a circular
to rectangular transition selects only one linear polarization. By
rotating the horns perpendicularly one with respect to the other,
we select two perpendicular linear polarizations $E_{0x}$ and $E_{0y}$
on the sky . If we denote with $E_{x}$ and $E_{y}$ the two orthogonal
components at the outputs of the horns, we can express them by \cite{key-7}: \begin{eqnarray}
E_{x} & = & \int d^{2}\vec{s}E_{0x}\left(\vec{s}\right)B\left(\vec{s}\right)e^{i\omega t}\label{eq:1}\end{eqnarray}
 and \begin{equation}
E_{y}=\int d^{2}\vec{s}E_{0y}\left(\vec{s}\right)B\left(\vec{s}\right)e^{i\left[\omega t+\phi_{path}(\vec{s})\right]}.\label{eq:2}\end{equation}

\noindent After that, $E_{x}$ and $E_{y}$ are twisted to the
same direction in order to be combined. Before being combined, a phase
shift $\Delta\alpha$ is introduced with a controllable phase shifter.
To simplify, we will consider that the phase shift is introduced in
only one of the channels : \begin{eqnarray}
E'_{x} & = & \int d^{2}\vec{s}E_{0x}\left(\vec{s}\right)B\left(\vec{s}\right)e^{i\omega t}\label{eq:1}\end{eqnarray}
 \begin{equation}
E'_{y}=\int d^{2}\vec{s}E_{0y}\left(\vec{s}\right)B\left(\vec{s}\right)e^{i\left(\omega t+\phi_{path}(\vec{s})+\Delta\alpha\right)}\label{eq:2}\end{equation}

Finally, the electromagnetic waves are combined in the 90\textdegree{}
hybrid coupler and the power received on each of the outputs over
the considered bandwidth is
\begin{equation}
P=\frac{1}{2Z_{w}}\intop EE^{*}d\nu\label{eq:P}
\end{equation}
where $Z_{w}$ is the wave impedance and $E$ the field resulting
from the combination of $E'_{x}$and $E'_{y}$. After calculation,
the detected power is given by the following expression:
\begin{eqnarray}
\left\langle \left|E\right|^{2}\right\rangle  & = & \frac{1}{2}\int B^{2}\left(\vec{s}\right)\left[I\left(\vec{s}\right)\pm U\left(\vec{s}\right)\sin\left(\phi_{path}+\Delta\alpha\right)\right]d^{2}\vec{s}\\
 & = & \frac{1}{2}\left[\int B^{2}\left(\vec{s}\right)I\left(\vec{s}\right)d^{2}\vec{s}\pm\left|V^{U}\right|sin\left(\phi_{U}+\Delta\alpha\right)\right]
 \label{eq:dibo}
 \end{eqnarray}
 where I and U are the Stokes parameters as defined in \cite{key-8}. The
visibility $V^{U}$ is the Fourier transform of the Stokes parameter
U on a field defined by the beam :\begin{equation}
V^{U}=\int{B\left(\vec{s}\right)^{2}U\left(\vec{s}\right)e^{-i2\pi\frac{\vec{D}.\vec{s}}{\lambda}}d^{2}\vec{s}}=\left|V^{U}\right|e^{i\phi_{U}}.\label{Vis}\end{equation}
 This visibility can be extracted from the measured signal by varying
the phase shifts \cite{key-9}. Using an OMT for each horn and a beam combiner with
eight ports (four inputs and four outputs), equation (\ref{eq:dibo}) can be generalized
and the visibilities of each 4 Stokes parameters are measured. All
the polarized information on the sky is therefore recovered and hence
the B-modes provided that the instrument has enough sensitivity and
systematic cleanness \cite{key-4, key-10}.

%\newpage{}

\section{System analysis}

An important issue to understand bolometric interferometry is systematic
effects extraction and laboratory characterization. In order to be
able to do that, we propose a method that allows to recover the system
parameters by a set of both power and vectorial measurements.

\noindent A crucial point is to link in a coherent way the wide bandwidth power measurements
and the monochromatic vectorial measurements. Using a VNA, we
measure the S parameters, while with the power meter (or bolometer)
we measure directly the power derived as a function of the incoming
EM field. The integration of the VNA measurements in a bandwidth $\Delta\nu$
makes it necessary to correctly simulate the behavior of the whole
system (calibration of the integration window). For that purpose we
consider the properties in term of S parameters for each component
of the system and we deduce the theoretical power at the output. Then
we show how to convert the properties encoded in the S-parameters
into a more general formalism that allows to take into account the
polarization states of the outgoing EM field.

\subsection{Analysis method}

There are two kinds of analysis we can use : two ports and four ports
analysis. Each component of the instrument, apart the hybrid coupler, can be studied independently
as a two ports device. With a VNA, we measure the corresponding {[}S{]}
matrix which can be related to the {[}ABCD{]} matrix. The \emph{ABCD}
formalism is a very useful tool for the quasi-optical instrument analysis
since it allows the propagation of the signal parameters linearly
over the chain \cite{key-11}. We use this analysis for all the two ports components
along each channel of the instrument since there is no mixing between
the incoming signals. However, this formalism does not take into account the two polarization
states of the incoming EM field and their combination.

From the definition of the S parameters,we can deduce a relation between the incoming
field $E_{k,1}$ and the outgoing field $E_{k,2}$ from a studied device k (see Appendix):
\begin{equation}
E_{k,2}=\left(\frac{S_{21}e^{-i\beta_{2}}+\left(\frac{S_{11}}{S_{12}}\right)e^{i\beta_{2}}}{e^{-i\beta_{1}}+S_{11}e^{i\beta_{1}}}\right)E_{k,1}
\label{eq:11}
\end{equation}
 where k$\in(x,y)$.

\noindent This equation shows the relation satisfied by a single element
of a Jones like matrix \cite{key-8}:

\begin{equation}
\left[\begin{array}{c}
E_{x2}\\
E_{y2}\end{array}\right]=\left[J\right]\times\left[\begin{array}{c}
E_{x1}\\
E_{y1}\end{array}\right]
\label{eq:12}
\end{equation}
where $E_{x1}$ and $E_{x2}$ are respectively the input and the output fields
in the x direction, $E_{y1}$ and $E_{y2}$ in the y direction and
$J$ a $2\times2$ matrix which can be written as:

\begin{equation}
\left[J\right]=\left[\begin{array}{cc}
J_{x} & \xi_{x}\\
\xi_{y} & J_{y}\end{array}\right]\label{eq:13}\end{equation}
In equation (\ref{eq:13}), $\xi_{x}$ and $\xi_{y}$ allow the superposition of the two
polarization states.

For a general four ports device or system, we will have $\xi_{x,y}\neq0$.
In order to extract the parameters of such a system, let's consider
a device with the incoming ports (1,2) and the outgoing ports (3,4).
In this case, $\xi_{x,y}$ expresses the transmission from port (1,2)
to port (4,3) when $J_{x,y}$ represents the transmission from port
(1,2) to port (3,4). With two matched loads at port (2) and (3), we
can find $\xi_{x}$ and using two matched loads at ports (1) and (4),
we can find $\xi_{y}$. With the matched loads at the ports (2) and
(4) we can recover $J_{x}$ and at the ports (1) and (3) we can recover
$J_{y}$. This allows us to reconstruct the signal in order to eliminate
the systematic effects.

For our instrument, this can be used for both the hybrid coupler and
the whole chain. In the output of the hybrid coupler, the power is
expressed by equation (\ref{eq:P}).

\section{Experimental setups}

Looking at equation (\ref{eq:dibo}), it is clear that there are two kinds of measurements
that can be done in order to demonstrate the basic operation of bolometric
interferometry. The first one consists in varying $\phi_{path}$ with
$\Delta\alpha$ fixed, while the other one is varying $\Delta\alpha$
with $\phi_{path}$ fixed. $\Delta\alpha$ is varied by changing the
phase introduced by the phase shifters and $\phi_{path}$ is changed
by rotating the source with respect to the phase center of the interferometer.
The bolometric measurements have been done varying $\phi_{path}$
and $\Delta\alpha$ while the room temperature measurements fixed $\phi_{path}$
changing $\Delta\alpha$. In the following, the two setups are described.\\

%\begin{figure}[t]
\begin{centering}
\includegraphics[bb=136bp 488bp 552bp 717bp,clip,width=8cm]{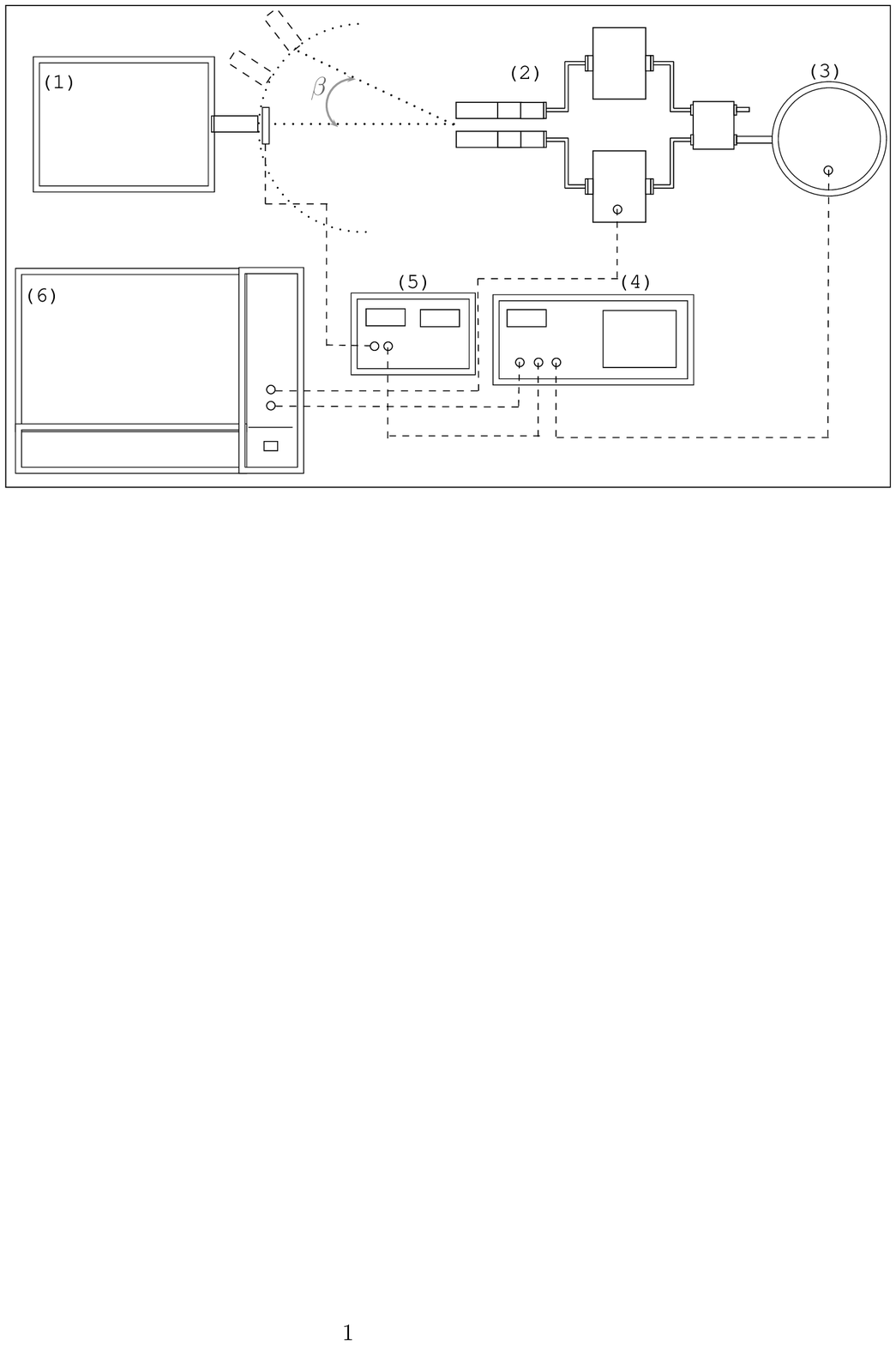}
\par\end{centering}
\caption{Experimental setup for bolometric measurements. (1) ABmm Vector Network Analyzer with the source polarized vertically, (2) The DIBO instrument, (3) Cryostat with the 4K bolometer, (4) Lock-in Amplifier, (5)Chopper, (6) Computer.\\}
\label{dibo_rot}
%\end{figure}
%\begin{figure}[t]
\begin{centering}
\includegraphics[clip,width=8cm]{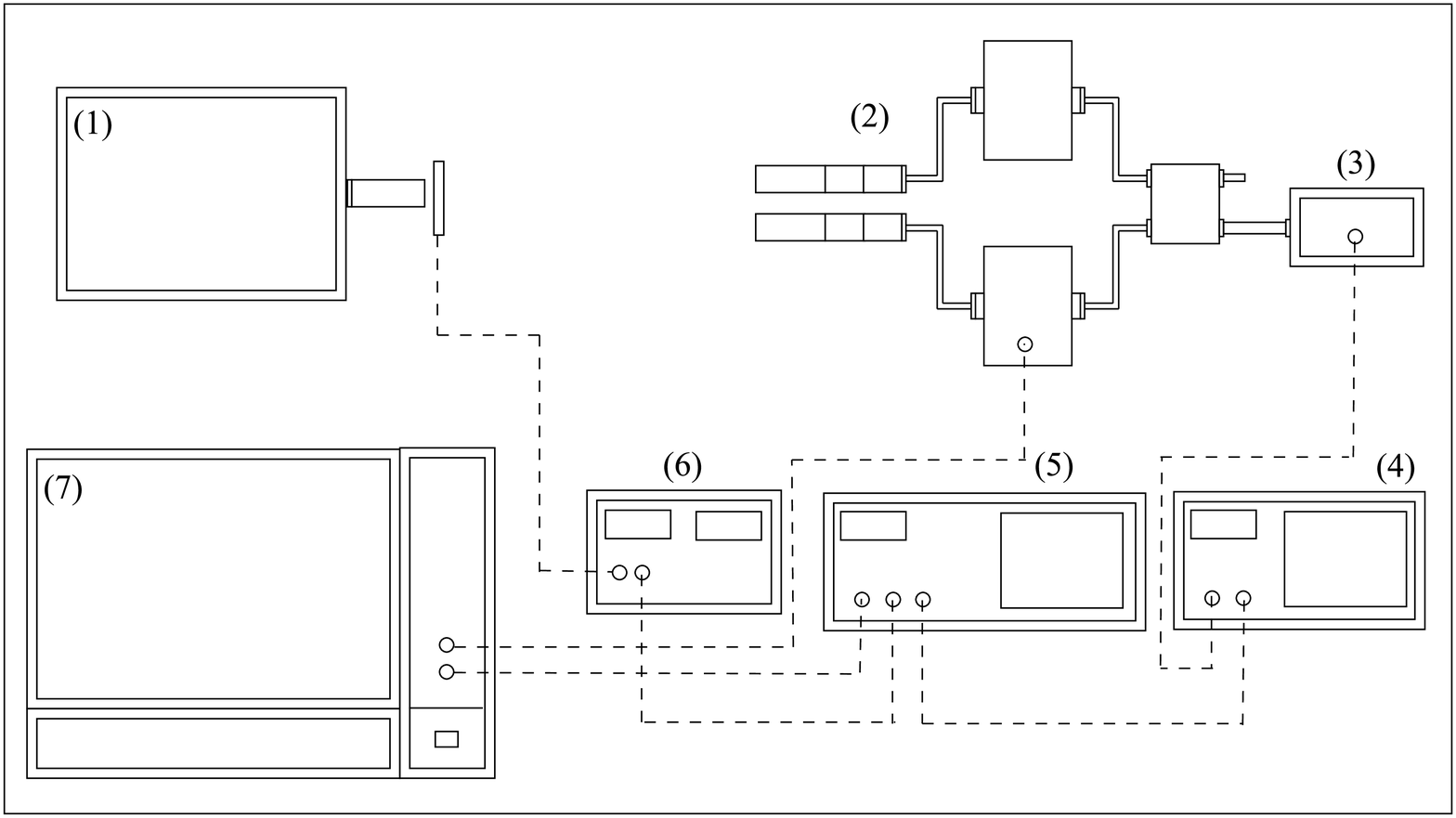}
\par\end{centering}
\caption{The set-up used for the characterization of the interferometer's response
to broad-band signals. (1) Farran BWO-10 power generator. (2) Demonstrator
to be characterized. (3) HP W8486A power sensor. (4) HP 438A power
meter. (5) EG\&G 5209 lockin amplifier. (6) Chopper controller. (7)
Computer for controllable phase variation and data acquisition.}
\label{pm_setup}
%\end{figure}

\subsection{Bolometric setup (4K)}

We used an ABmm MVNA%
\footnote{http://www.abmillimetre.com/%
} as a monochromatic source and a 4 K cooled semiconductor bolometer%
\footnote{IR lab%
} read by a lock-in amplifier as a detector (figure \ref{dibo_rot}).

In order to change $\phi_{path}$, we move the instrument with respect
to the source by an angle $\beta$ as shown in figure \ref{dibo_rot}. The distance
between the two horns of the instrument is set to $20.5$ cm and the
gaussian beam of the emitting horn has a 1$\sigma$ width of $\sim4.2$\textdegree at 90 GHz.

\subsection{Room temperature setup}\label{RoomTs}

In this section the experimental set-up used for power measurements
around the nominal central frequency of DIBO (90 GHz) is described (figure \ref{pm_setup}).
A frequency sweeper FARRAN BWO-10 is used as a source, operating both
in Continuous Wave (CW) mode and in sweep mode. In the first case,
a quasi-monochromatic signal is produced at a fixed frequency. In
the second case, a frequency sweep across a chosen bandwidth (87-93
GHz, or 75-110 GHz in our case) is performed in 10 ms. The signals,
collected by the two antennas of the interferometer, after phase shifting
are combined in the hybrid, and then detected with a power sensor
HP W8486A, placed on one of the two outputs of the combiner (the second
one is closed on a matched load). Then, the analog output of the power
meter HP 438A feeds a single channel lock-in amplifier EG\&G 5209.
The external reference of the lock-in is the driving signal coming
from the control unit of the chopper, which modulates the source signal
at 10 Hz. The basic idea behind this set-up is to replace an artificial
black body with a fast frequency sweep across all the frequency band.
If the sweep is much faster than the response time of the power meter
(as in our case, since the typical time constant of the power sensor
is $\tau$ =35 ms), this system can mimic a broad-band source. Of
course the best solution would be an artificial black-body, but commercial
power meters operating at room temperature are not sensitive enough
to detect a 300- 500 K signal. This set-up can be useful for the characterization
of millimeter-waves systems whenever cryogenic front-ends are not
available.

We underline that even if the signal of the sweeper is strong enough
to be directly detected by the power sensor, we preferred to use a
lock-in in order to operate always with a favorable signal-to-noise
ratio, even when the source's oscillator was used at power levels
below the 25$\%$ of the peak power available. In fact below this
threshold the sweeper's output power \textit{vs} frequency is practically
flat. For our purposes, using the instrument's specifications, we
can estimate that $P_{83-97}/P_{75-110}\cong\Delta\nu_{83-97}/\Delta\nu_{75-110}=0.17$:
that is, the power delivered in a frequency range is essentially proportional
to the bandwidth itself.

%\newpage{}

\section{Results}

\subsection{Characterization}

The S-matrix of every single microwave component has been obtained
before assembling the full instrument. Then, the full interferometric
chain, from the twist to the output ports of the hybrid, has been
characterized in the same way, allowing us to get the transmission
function of the full instrument. In such a system, the devices deserving
the most careful characterization are phase shifters. In fact, commercial
phase shifters behaves usually in a nearly ideal way when they are
operated close to the center of the nominal frequency range, while some
unwanted effects start to appear at other frequencies. In particular,
we noticed that, going far from 90 GHz, (1) losses become increasingly
important and (2) changing the phase settings produces an amplitude
modulation of the incoming signal (see figure \ref{fig:phase}). Of course the point here is only
to know and correctly recognize such effects, and eventually see how
they propagate till the output of the system. At 90 GHz, the nominal
center frequency, the power transmission coefficient is almost flat
while the nominal phase setting of the device varies between 0 and
360 degrees: the values of $\left|S_{21}\right|$ in linear scale
oscillates between 0.81 and 0.82. At 110 Ghz, the same quantity is
modulated between 0.36 and 0.84, while at 80 GHz between 0.77 and
0.87 (the same overall behavior and the same values are found measuring
$\left|S_{12}\right|$, the device being reciprocal).

%\begin{figure}[t]
\begin{centering}
\includegraphics[clip,width=7cm]{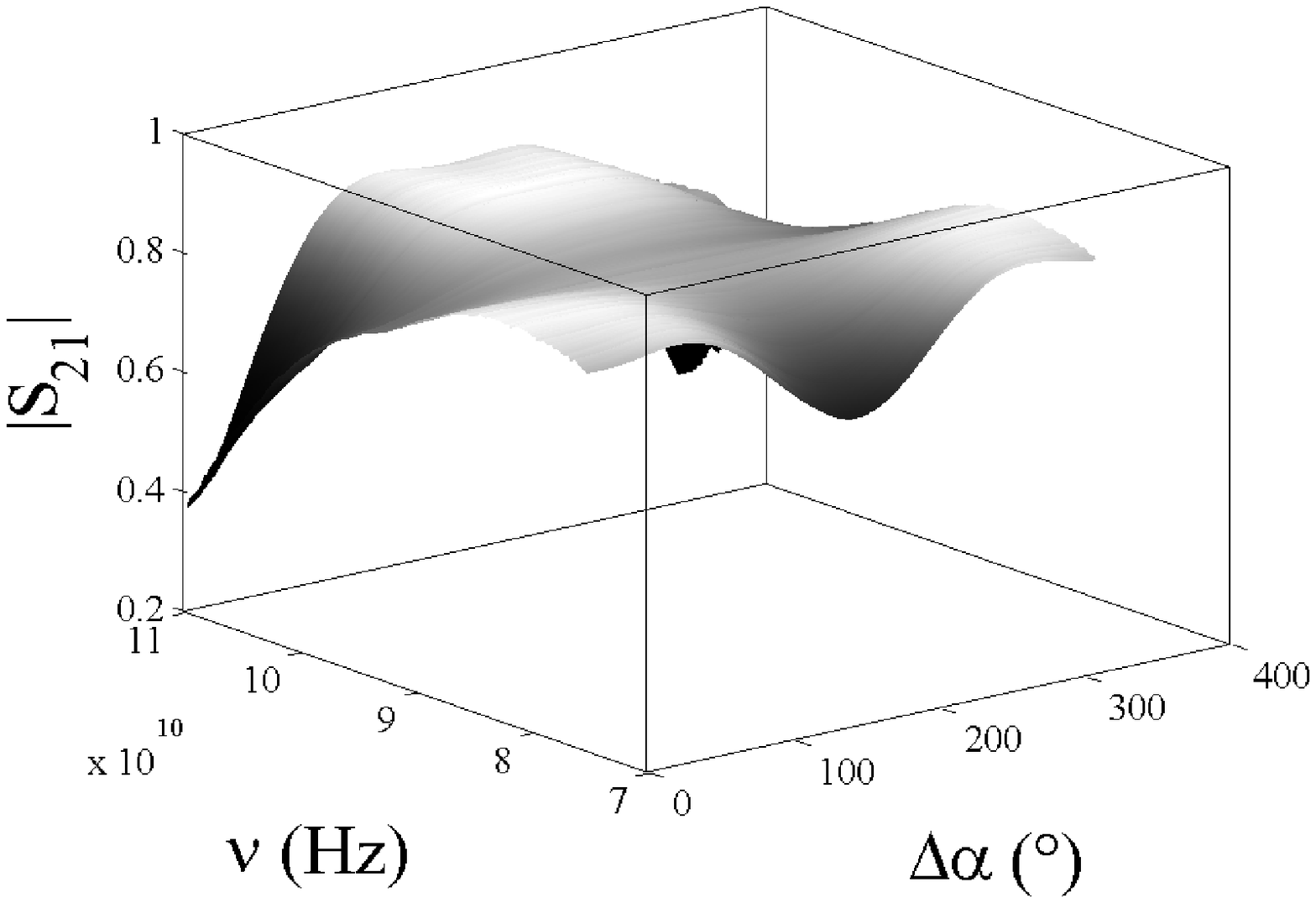}\includegraphics[clip,width=9cm]{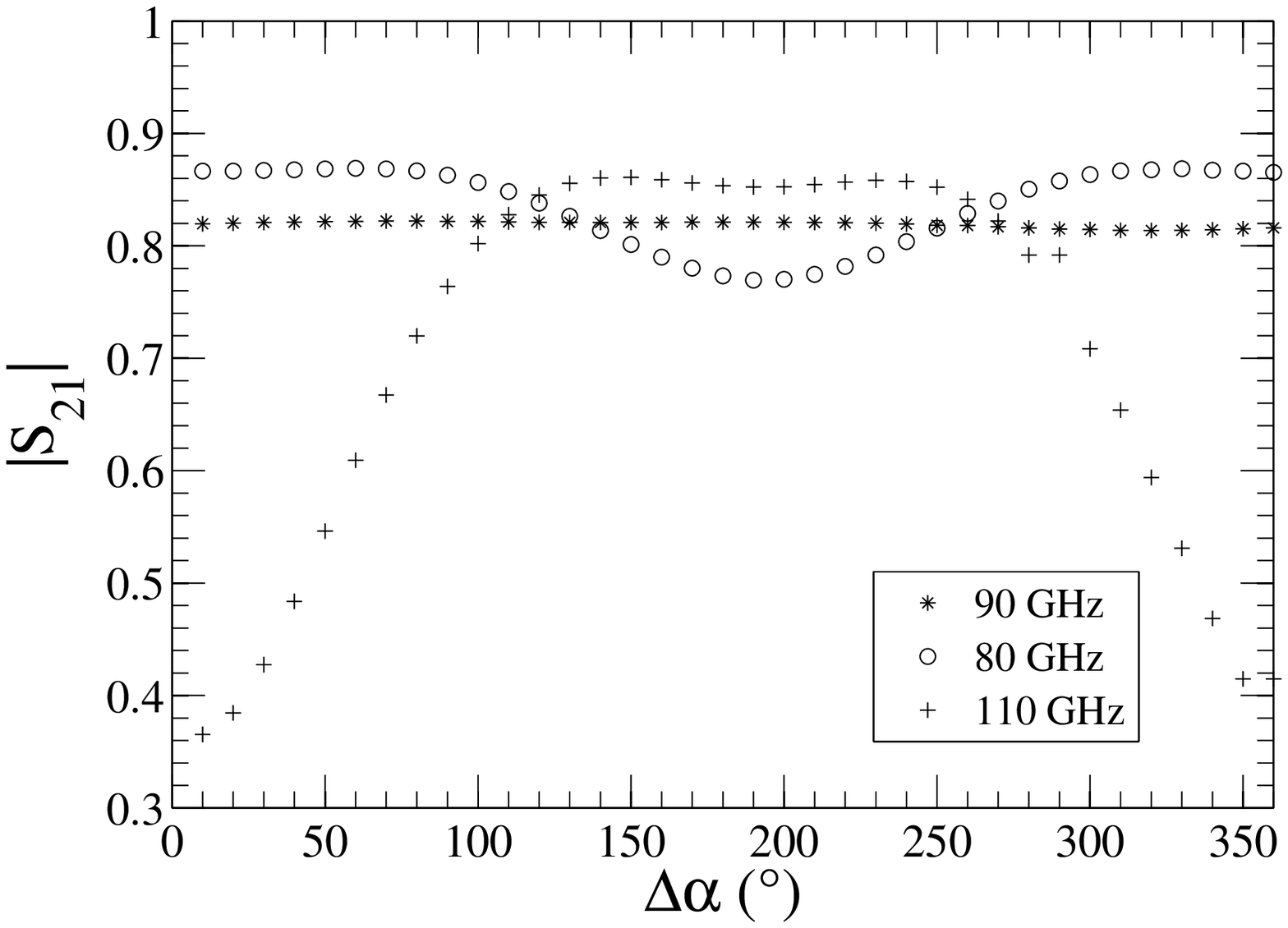}
\par\end{centering}
\caption{Right : $|S_{21}|$, for one phase shifter, as a function of $\Delta\alpha$ at 80, 90 and 110 GHz. Left : $|S_{21}|$ as a function of $\nu$ and $\Delta\alpha$.\\}
\label{fig:phase}
%\end{figure}
We measured the behavior of the full chain at each frequency as a
function of the phase introduced by the phase shifter with a VNA.
We report here the received power as a function of the introduced
phase shift. This power is normalized with respect to the ideal behavior
of the instrument and represents the monochromatic transmission function.
As we can see in figure \ref{fig:normalised power}, the instrument is designed to work at 90
GHz and its efficiency decreases when we switch to another frequency
in the bandwidth 75-110 GHz. This is due to the fact that the devices
are designed to work only close to the center frequency band. Moreover,
we can observe that the amplitude does not depend only on the frequency
but also on the phase. This effect is mainly due to the phase shifter.
\begin{center}
%\begin{figure}[t]
\begin{centering}
\includegraphics[width=10cm]{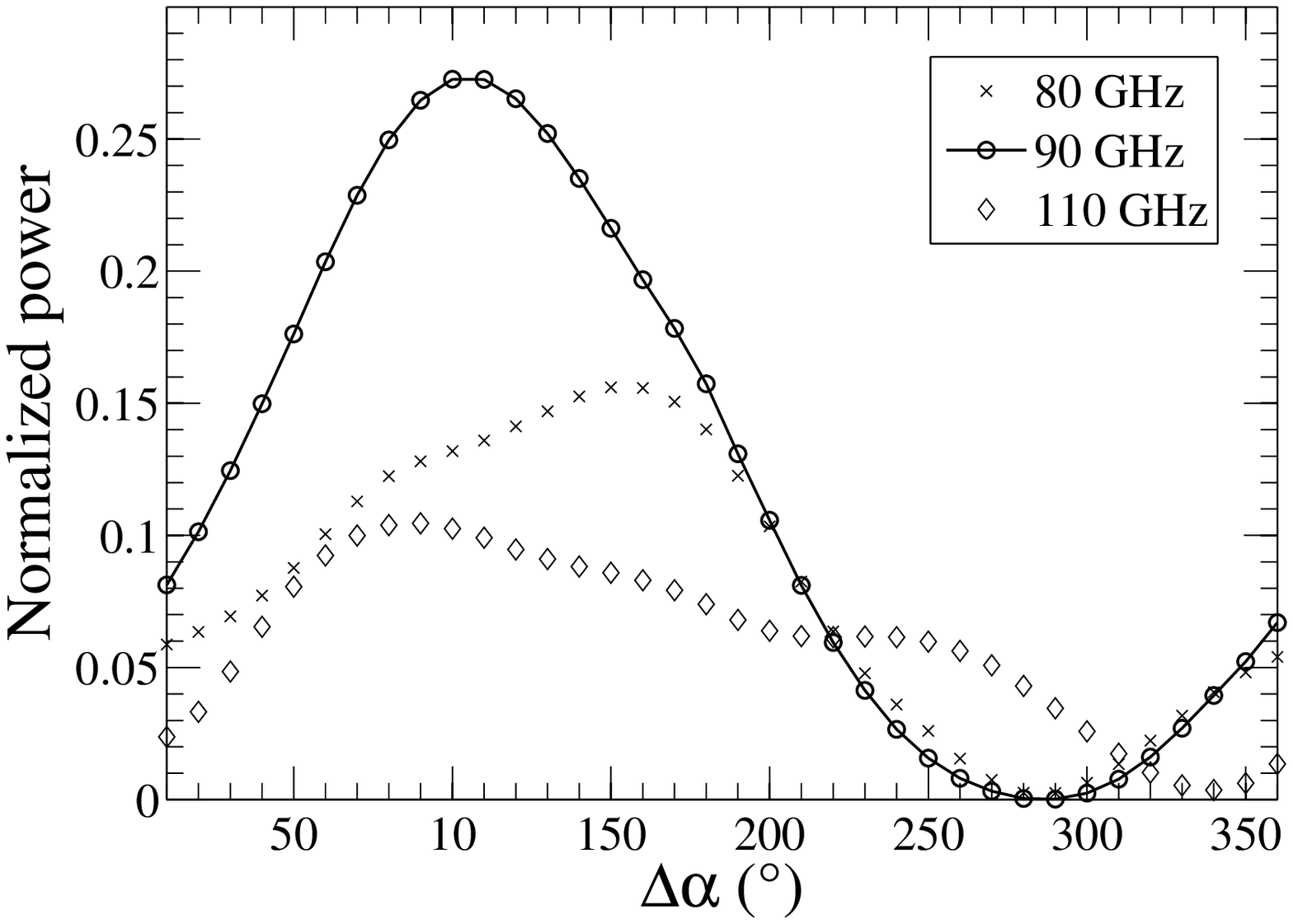}
\par
\end{centering}
\caption{Normalized output power as a function of the phase shift (phase shifters) for three frequencies 80, 90, 110 GHz.}
\label{fig:normalised power}
%\end{figure}
\par\end{center}

\subsection{Bolometric measurements}
%\begin{figure}[t]
\begin{centering}
\includegraphics[clip,width=10cm]{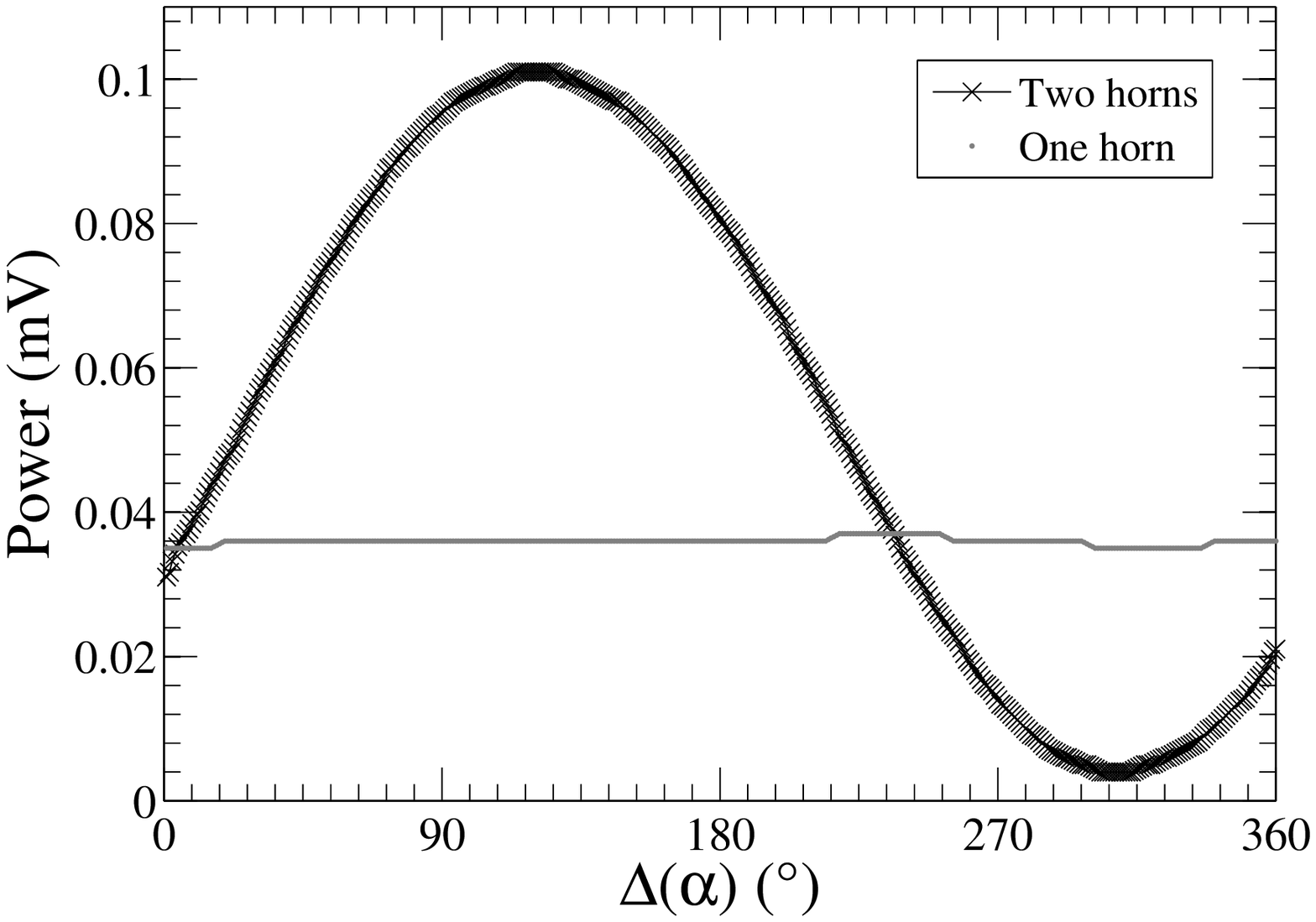}
\par\end{centering}
\caption{Measured signal from the bolometer with two horns (solid line) and hiding one of the horns (dot line).\\}
\label{dibo90}
%\end{figure}
Figure \ref{dibo90} shows bolometric measurements at 90 GHz by varying the phase
shift $\Delta\alpha$ introduced by the phase shifters, $\beta$ being fixed. As expected from equation (\ref{eq:dibo}),
we observe a sinusoidal signal on the bolometer. Hiding one of
the two horns, the observed interference disappears. The observed
amplitude modulation is therefore produced by the interference and
not by an amplitude modulation due to the phase shifters.

 As shown on figure \ref{dibo_rot}, we rotate the source with respect to the instrument by an angle $\beta$. We can therefore calculate from equation (\ref{eq:dibo}) the power at the output of the hybrid coupler as a function of the parameters of the setup:

\begin{equation}
P\left(\beta\right)=UB^{2}\left\{ 1+\sin\left[2\pi f\sin\left(\beta-\beta_{0}\right)+\theta\right]\right\} \label{eq:Pb}
\end{equation}
where $f=\frac{D}{\lambda}$, $D$ being  the distance between the horns
and $\beta_{0}$ an offset angle. $\theta$ is a factor that includes
the phase shift $\Delta\alpha$ introduced by the phase shifters and
other frequency dependent effects. These effects can be due to the
misalignment of the horns, the length difference of the channels and
the non ideal behavior of the hybrid coupler and the phase shifters
(Figure \ref{fig:phase}). U is the Stokes parameter of the source assumed to be a
point source. In the case of a gaussian profile of the beam, we have:
$B^{2}\left(\beta\right)=\exp\left(\frac{-\left(\beta-\beta_{0}\right)^{2}}{2\sigma^{2}}\right)$.\\

%\begin{figure}[t]
\begin{centering}
\includegraphics[bb=0bp 70bp 948bp 425bp,clip,width=12cm]{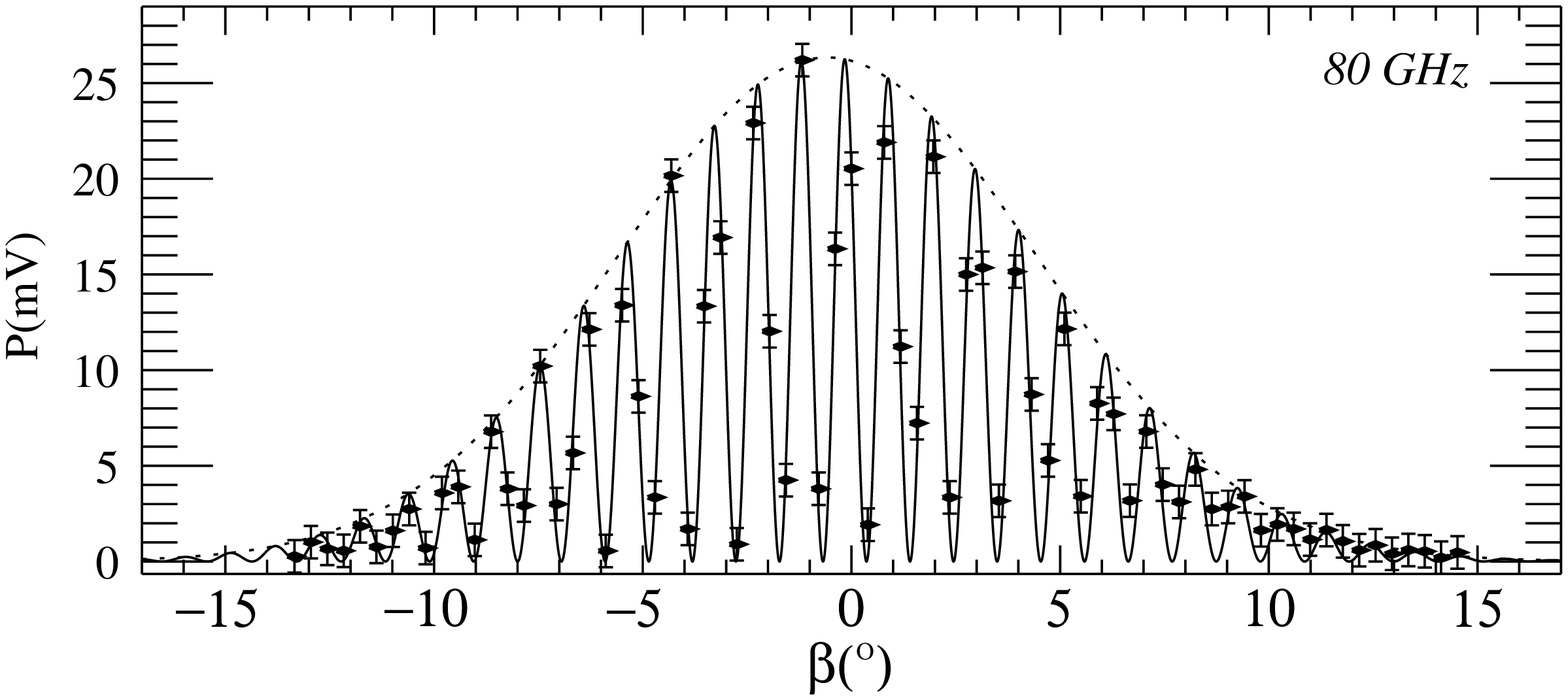}
\includegraphics[bb=0bp 70bp 948bp 425bp,clip,width=12cm]{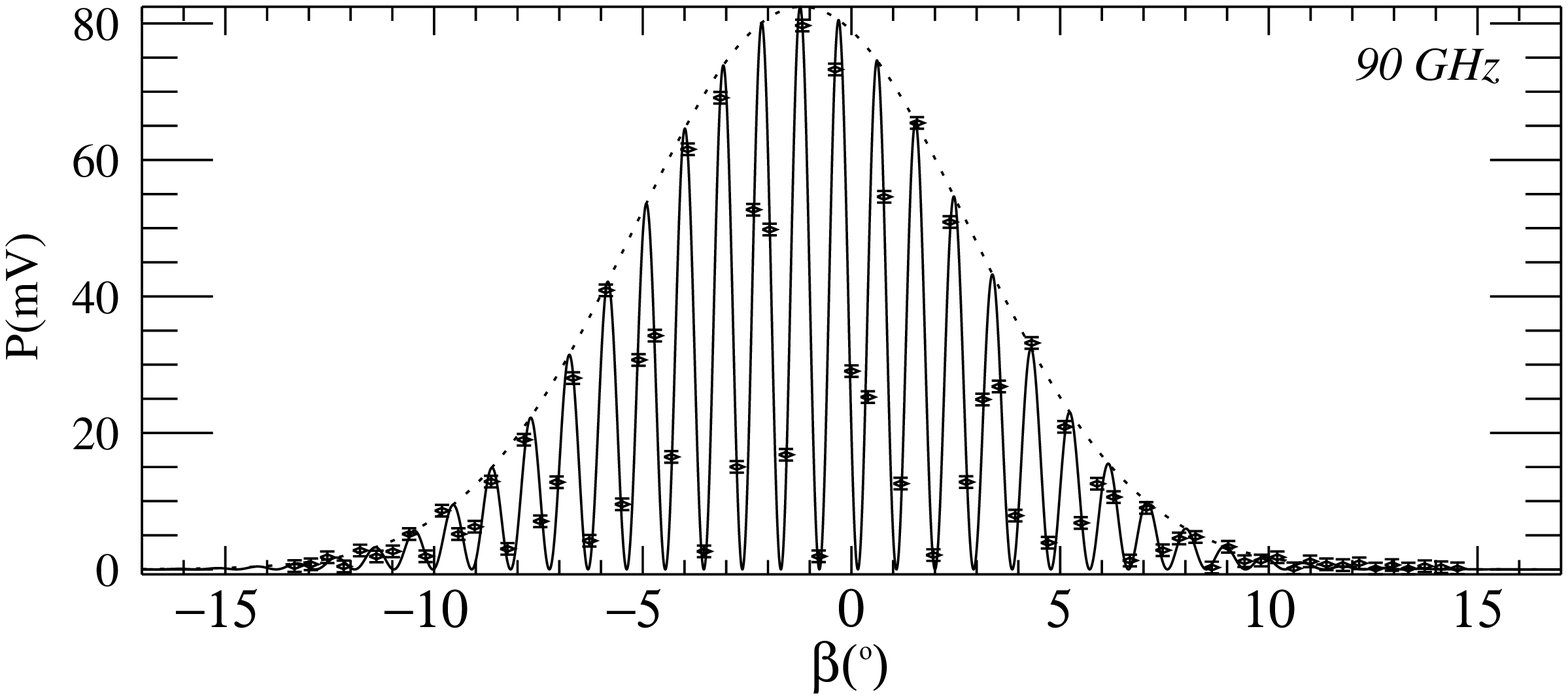}
\par\end{centering}
\caption{Measured signal (diamonds) and fitted fringes (solid line) as a function of the rotation angle with respect to the source at two frequencies (90 GHz and 80 GHz). The dashed line represents the gaussian envelope.\\}
\label{fringes}
%\end{figure}

In figure \ref{fringes} , we represent the measured power as a function of $\beta$
at $80$ GHz and $90$ GHz. We also represent the fit of the equation
(\ref{eq:Pb}) in the same figure. The values of the fitted parameters are showed
in the table \ref{tab:Tf}. The expected fringe frequency f is $54.7\pm0.5$ for
$80GHz$ and $61.5\pm0.6$ for $90GHz$. We can see from table \ref{tab:Tf}
that the expected values of $f$ and $\sigma$ ($\sim4.2$\textdegree at 90 GHz) are close to the parameters
extracted from the fit. The figure \ref{fringes}, therefore, shows the interference
patterns from the two horns shaped by the beam.\\

\begin{table}[H]
\begin{centering}
\begin{tabular}{ccc}
\toprule
$\nu$ & $80$ GHz & $90$GHz\tabularnewline
\midrule
\midrule
$U$ (mV) & $13.17\pm0.179$ & $41.225\pm0.199$\tabularnewline
\midrule
$\beta_0$ (\textdegree) & $-0.58\pm0.078$ & $-1.18\pm0.022$\tabularnewline
\midrule
$\sigma$ (\textdegree) & $5.01\pm0.079$ & $4.019\pm0.022$\tabularnewline
\midrule
$f$ & $54.97\pm0.049$ & $62.101\pm0.021$\tabularnewline
\midrule
$\theta$ (\textdegree) & $305.38\pm27.065$ & $106.75\pm8.754$\tabularnewline
\bottomrule
\end{tabular}
\par\end{centering}
\caption{Fit parameters of the equation \ref{eq:Pb} with $\chi^2=15.78$ for 90 GHz and $\chi^2=1.99$ for 80 GHz.}
\label{tab:Tf}
\end{table}

\subsection{Room temperature measurements}

As described in \ref{RoomTs}, the results showed here have been done with a system that mimics a broadband source. Figure \ref{Power6} and \ref{PowerW} represent the power response of the system in arbitrary units. We can see then clearly interference patterns.

%\begin{figure}[t]
\begin{centering}
\includegraphics[bb=175bp 50bp 700bp 450bp,clip,width=8cm]{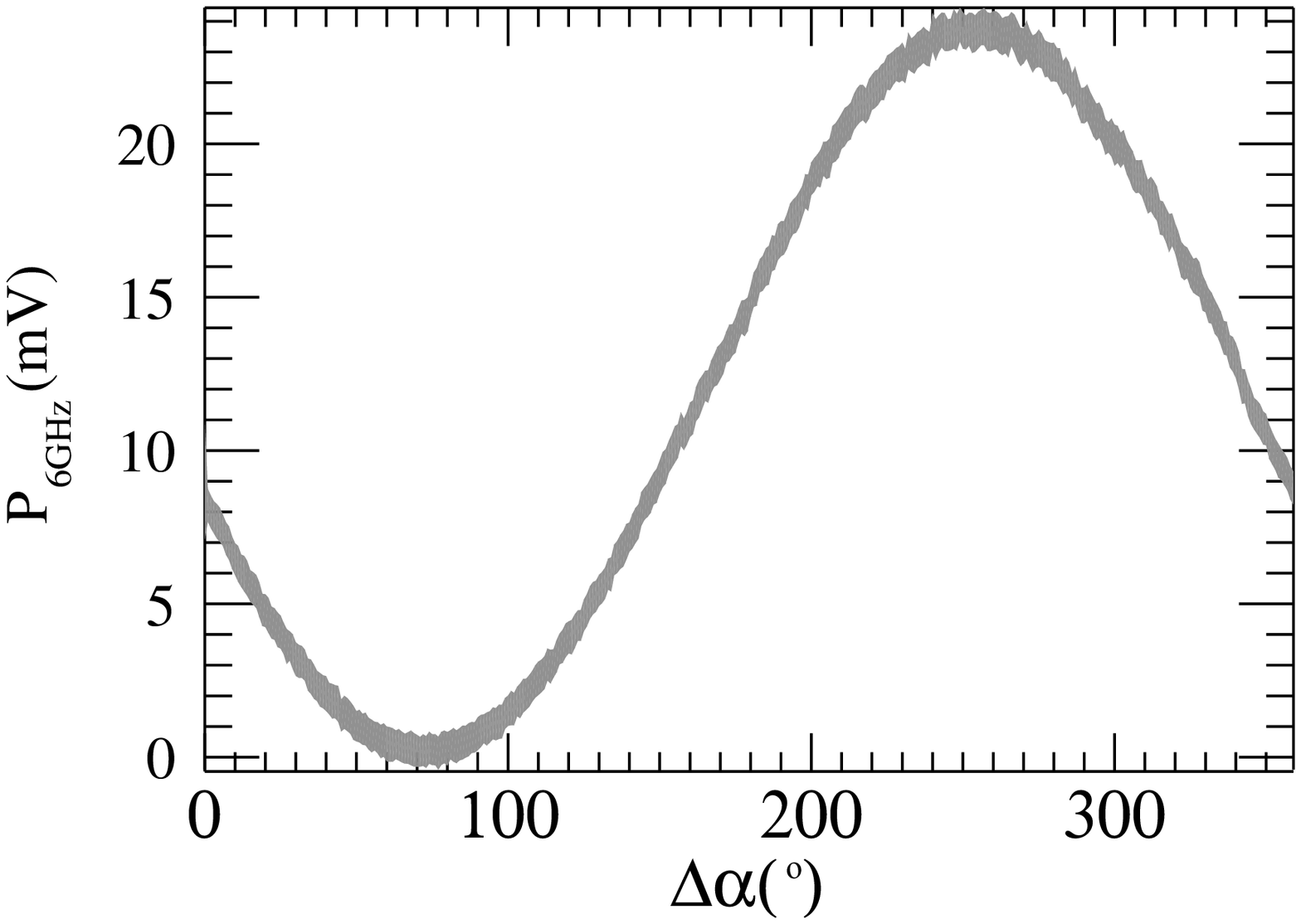}\includegraphics[bb=175bp 50bp 700bp 450bp,clip,width=8cm]{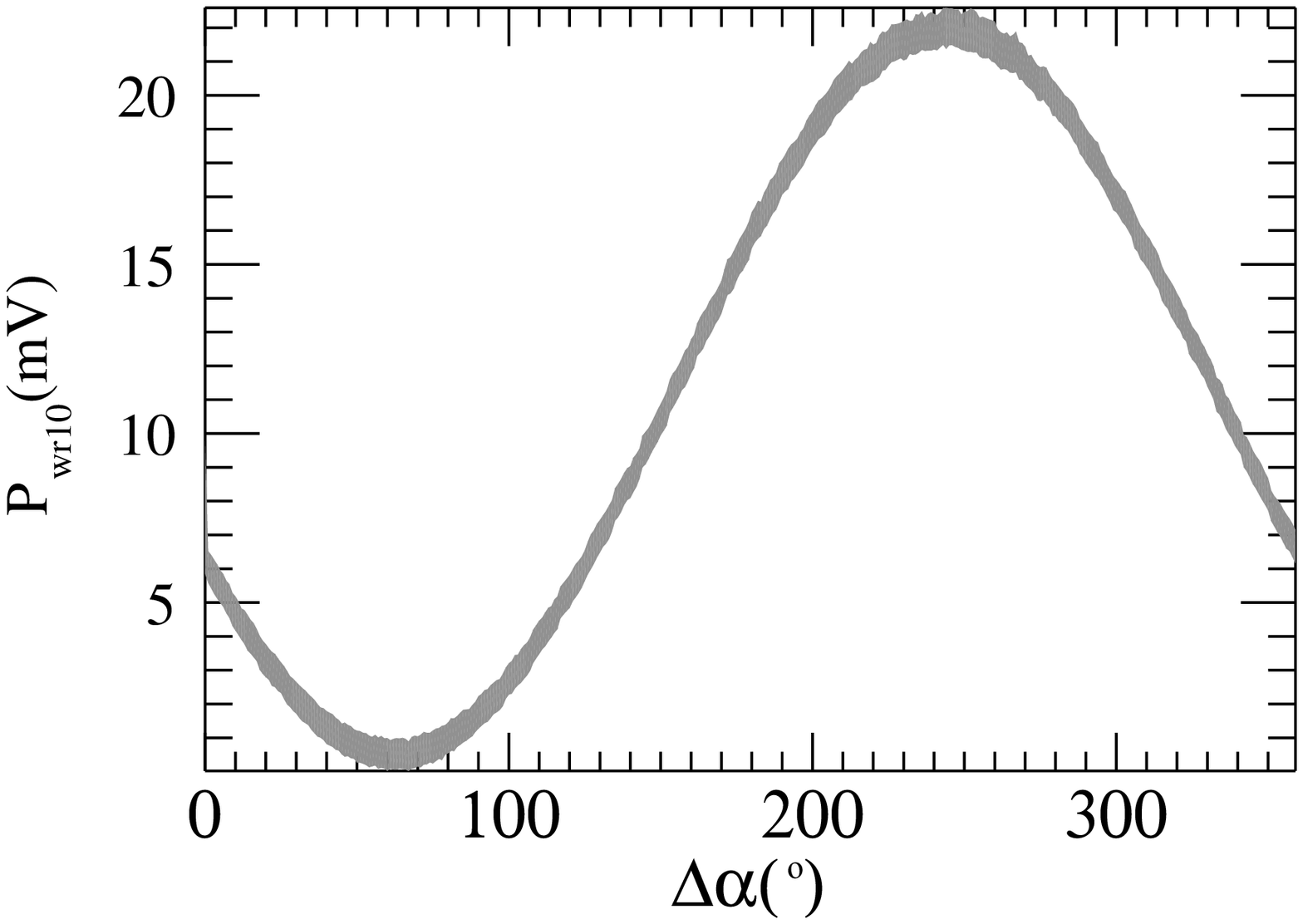}
\par\end{centering}
\caption{(a) Left: measured power with the room temperature bench in $6$ GHz bandwidth around $90$ GHz. (b) Right: measured power with the room temperature bench in the W band (between $75$ GHz and $110$ GHz).}
\label{Power6}
%\end{figure}
%\begin{figure}[t]
%\begin{centering}
%\par\end{centering}
%\caption{Measured power with the room temperature bench in the W band (between $75$ GHz and $110$ GHz)}
%\label{PowerW}
%\end{figure}

\subsubsection{Data analysis}

\subsubsection*{\textmd{\textit{Calibration factor}}}

When performing measurements on a given bandwidth $\Delta\nu$, the
system output is affected by the effect of bandwidth smearing. This
effect alters the fringes frequency and compress their dynamic range,
reducing their height. In order to verify the correctness of the power
measurements and to try to disentangle the smearing effect from the
systematics imprint, we have developed a strategy to compare power measurements with the
monochromatic VNA ones. As a starting point, having the DIBO full
chain transfer function and, starting from equation (\ref{eq:P}) and (\ref{eq:12}), we can reconstruct
the transmitted power as a function of the frequency and the phase
shift
\begin{equation}
T_{power,i}^{VNA}(\nu,\Delta\phi)=\left(J_{i}+\xi_{i}\right)\left(J_{i}+\xi_{i}\right)^{\star}
\label{eq:11}
\end{equation}
 where $i=x,y$.

\noindent The link with the power measurements is made doing a very
narrow band (CW) measurement at a frequency of 90 GHz to be compared with
the transmitted power $T_{power}^{VNA}$; comparing the correlated part of these
measurements we have:
\begin{equation}
P(90GHz,\Delta\phi)=P_{inc}T_{power}^{VNA}(90GHz,\Delta\phi)
\label{eq:26}
\end{equation}
Through this relation we extract $P_{inc}$, the fractional power
radiated by the sweeper and detected by DIBO. This parameter will
be used as calibration factor for all the other power measurements. A transmitted power function can then be generated for finite bandwidth by:
\begin{equation}
T_{power}(\Delta\nu,\Delta\phi)=\frac{P(\Delta\nu,\Delta\phi)}{P_{inc}}
\label{eq:Tp}
\end{equation}
Numerically integrating $T_{power}^{VNA}$ in the proper bandwidth
range, we derive the expected power output starting from VNA data:
\begin{equation}
T_{power}^{VNA}\left(\Delta\nu,\Delta\phi\right)=\int_{\Delta\nu}T_{power}^{VNA}\left(\nu,\Delta\phi\right)d\nu\
\end{equation}
$T_{power}(\Delta\nu,\Delta\phi)$  and $T_{power}^{VNA}\left(\Delta\nu,\Delta\phi\right)$ are normalized with respect to the perfect behavior of the instrument before being compared.

\subsubsection*{\textmd{\textit{Normalization}}}

Recalling equation (\ref{eq:12}), the signal detected in the power meter sensor
is the frequency integration of the square modulus of the linear combination of the incoming fields
in the output of the hybrid. In an ideal case the transmission function
has a rectangular shape so that the sinusoidal monochromatic response
of the system is multiplied by a $sinc$ function that depends on the
bandwidth to give the integrated response
\begin{equation}
P=P_{0}\left[1\pm\sin\left(\frac{\pi\delta}{c}\left(\nu_{1}+\nu_{2}\right)\right)\frac{\sin\left(\frac{\pi\delta}{c}\Delta\nu\right)}{\frac{\pi\delta}{c}\Delta\nu}\right]
\label{eq:}
\end{equation}
where $\delta$ is the optical path difference, $\nu_{1}$ and $\nu_{2}$
the frequency boundaries, $c$ the speed of light and $\Delta\nu$
the frequency bandwidth.

In the real case, we include the effect of the real transmission function
calculated with the full chain VNA measurements. This transmission
function is integrated in frequency with a numerical routine,
to find the expected integrated behavior. In order to correctly
normalize the results, we integrated also the ideal case response and
normalized it to the expected value. This normalization
factor is then applied to the real case integrated signal. Figures \ref{Power6v1} and \ref{PowerWv1} show the comparison between the expected and the measured signal for both a 6 GHz band centered at 90 GHz and the full WR10
band respectively.

\subsubsection*{\textmd{\textit{Complex degree of coherence}}}

\noindent As we can see in table \ref{vis}, there is a discrepancy between the expected and the measured
data. To check if it is something due to the applied normalization
we used, we calculate the modulus of the complex degree of coherence
for both the simulated and the measured signals, according to its
definition \cite{key-12}:
\begin{equation}
\left|\gamma\right|=\frac{I_{max}-I_{min}}{I_{max}+I_{min}}
\label{eq:}
\end{equation}

\noindent It turns out that the degrees of coherence for the simulated signals
is comparable with those of the measured ones (at least within the
error-bars), suggesting that the discrepancy is due to the normalization
we introduce.

\begin{table}[H]
\begin{centering}
\begin{tabular}{ccc}
\toprule
$\left|\gamma\right|$ & Measured & Simulated\tabularnewline
\midrule
\midrule
6 GHz Band & 0.96$\pm$0.04 & 0.94\tabularnewline
\midrule
WR10 Band  & 0.99$\pm$0.04 & 0.98\tabularnewline
\bottomrule
\end{tabular}
\par\end{centering}
\caption{Modulus of the complex degree of coherence computed for both the measured and simulated band integrated responses.}
\label{vis}
\end{table}

\subsubsection*{\textmd{\textit{Systematic effects}}}

\noindent A direct extraction of most of the systematics from the power signal
cannot be performed since the error bars are greater than any of the
imprint expected from them.

\noindent Anyway, some of the effects unveiled by the VNA characterization such
as the amplitude distortion induced by the phase shifter are not seen
because of the combined effect of the frequency integration of the
signal and the low transmission of the instrument. That is, amplitude
distortion is observed far from the central bandwidth frequency, where
the transmission is very low ($<-15$ dB) so that their relative contribution
to the output in the integration is negligible. This is confirmed
by the fact that, in the numerical integration of the VNA (monochromatic)
data, we do not observe these distortions. At some extent the frequency
integration of the signal washes out systematics and this is particularly
true for those affecting the signal far from the central frequency.
This effect, acting also as a transmission reduction, mimics the shrinking
of the effective bandwidth.

\noindent Finally, the normalization technique we used to compare
simulated and measured data is based on the assumption that the calibration
factor is the same at each frequency. If this is not true the frequency
dependance of the calibration can introduce a a different weighting
in the integration, inducing distortions.

%\begin{figure}[t]
\begin{centering}
\includegraphics[bb=50bp 10bp 1000bp 650bp,clip,width=9cm]{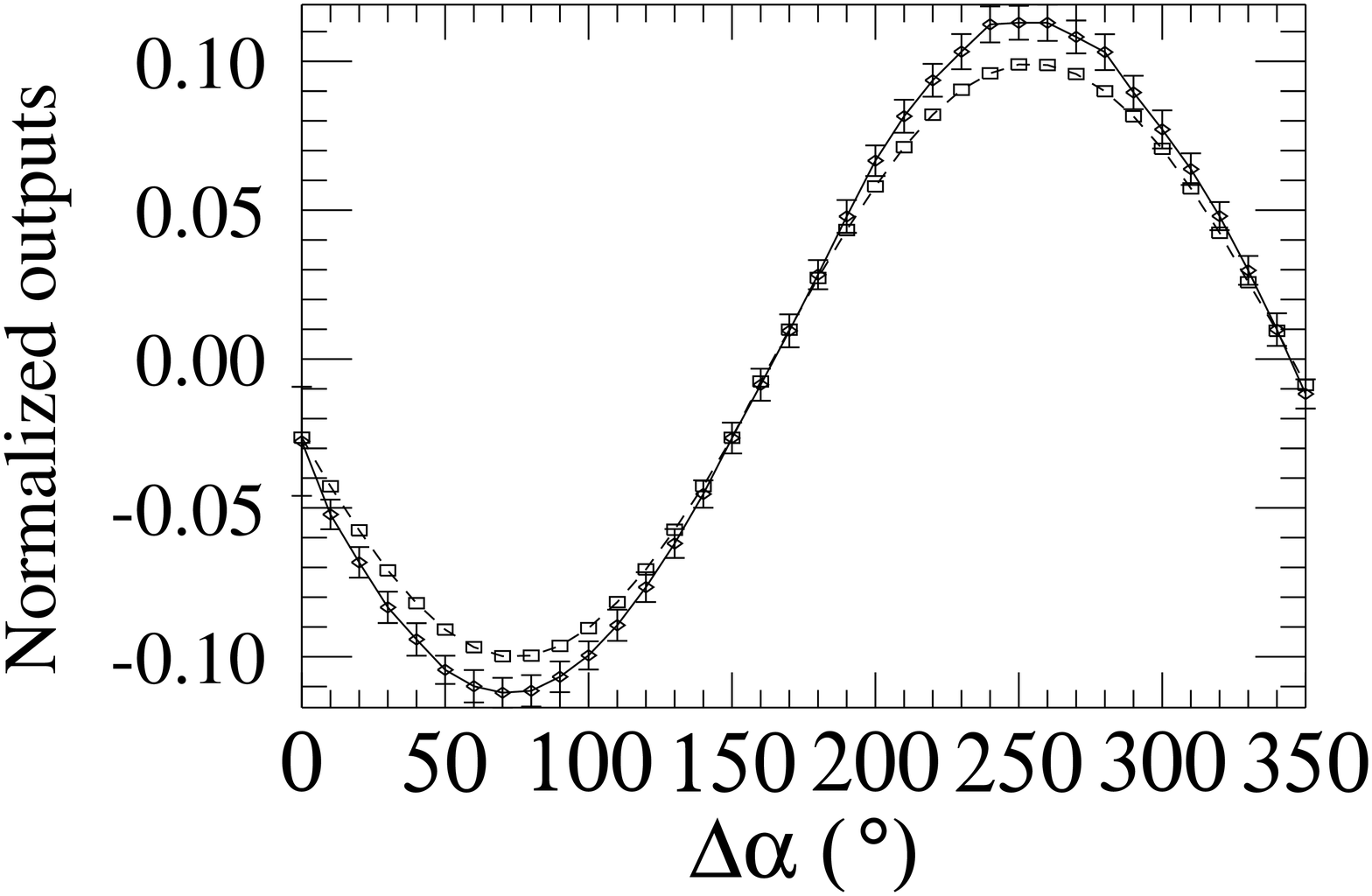}
\par\end{centering}
\caption{Comparison between the normalized correlated part of DIBO output signal (solid line) and the simulated one obtained integrating the VNA measurements (dashed line) for a 6 GHz band wide stimulus centered at 90 GHz.}
\label{Power6v1}
%\end{figure}
%\begin{figure}[t]
\begin{centering}
\includegraphics[bb=50bp 10bp 1000bp 650bp,clip,width=9cm]{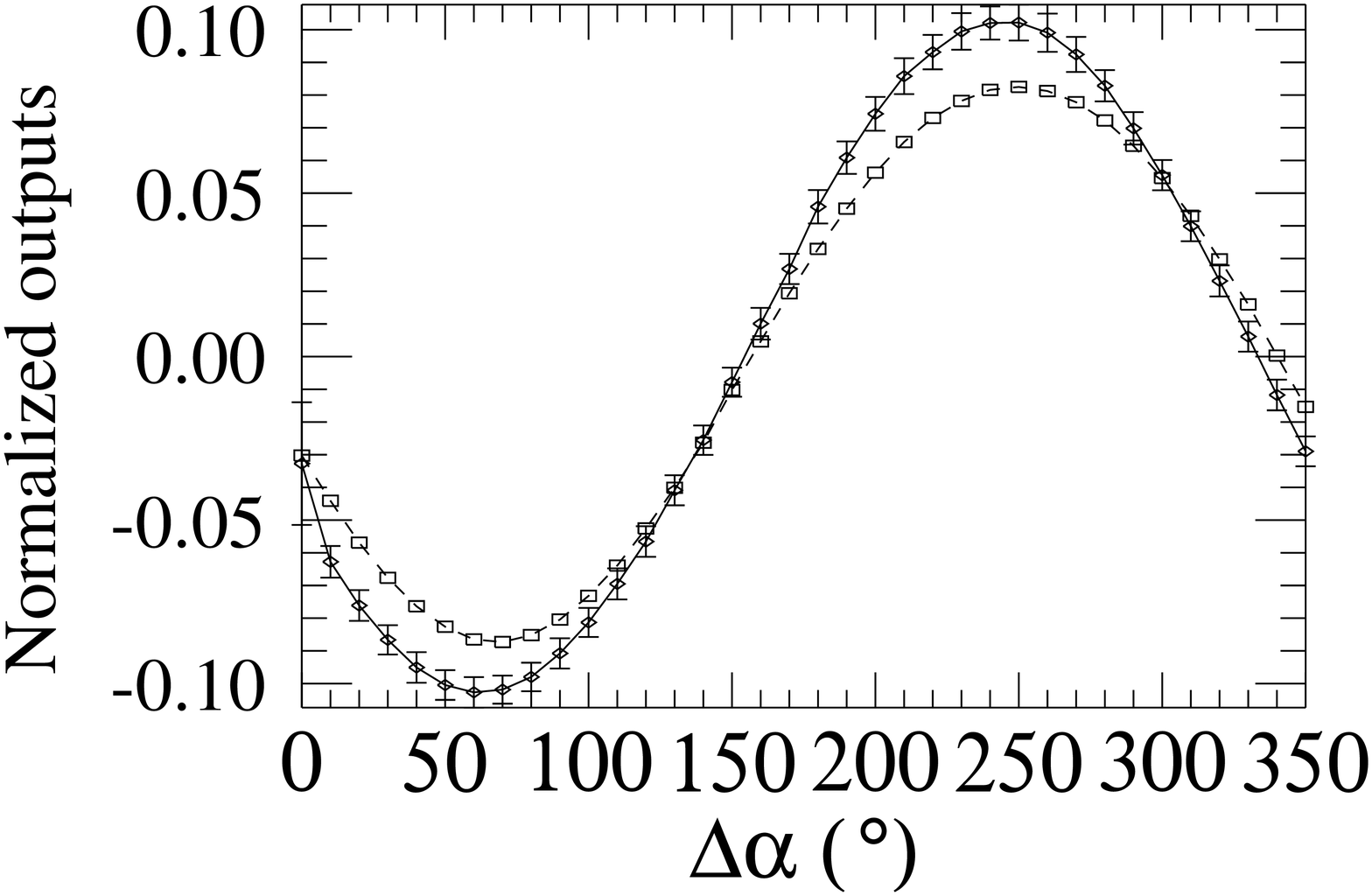}
\par\end{centering}
\caption{Comparison between the normalized correlated part of DIBO output signal (solid line) and the simulated one obtained integrating the VNA measurements (dashed line) for the full WR10 band wide stimulus.}
\label{PowerWv1}
%\end{figure}

%\newpage{}

\section{Conclusion and future perspectives}

With this demonstrator and a 4 K bolometer, the principle
of bolometric interferometry has been proved with success in a laboratory
environment.

This is the first attempt to understand how signals (their polarization
properties) are altered propagating along the interferometer branches
and how they interfere over a broad frequency band. In such a way we can study both the propagation of
instrumental systematics and the effect of bandwidth smearing. To
do so, we developed an experimental set-up that we used to illuminate the interferometer over the WR10 frequency band,
exploiting a fast sweeper able to cover the 75-110 GHz band in a fraction
of the time constant of the commercial power meter used as detector.
In this way, we have been able to check the very basic properties
of the response of the full instrument at 300 K. In the second place,
we have suggested a way of connecting the output of VNA measurements,
the scattering parameters, first to the Jones matrices, and then to
the modulated output power of the interferometer. The last one is
an important point since it provides a link between the immediate
output of microwave network analyzers, the S-matrix, and the elements
of the Jones matrices often used by cosmologists to propagate instrumental
systematics into astrophysical observables.

Even more important, illuminating
the interferometer with an artificial source and detecting the signal
with a 4 K bolometer, interference fringes have been clearly measured
at two different frequencies (80 and 90 GHz). In this way, we could
recover the expected fringe frequency.

In principle, a DIBO like architecture can be extended to a larger
number of baselines and used in combination with other technologies.
Indeed, to build a real bolometric interferometer (BRAIN/MBI), planar
superconductor components (filters, phase shifters and detectors)
will be considered as well as a beam combiner for the full power combination.
It's worth underlining that DIBO is not an instrument designed to
detect astronomical sources, but rather strong laboratory signals.

A future step will be the design of a more efficient demonstrator,
endowed with more sensitive detectors, able to observe both polarized
and unpolarized sources (artificial black-bodies). Then, having a
more sensitive demonstrator, we can also test the capabilities of
our conceptual scheme in unveiling systematics hidden in the modulated
output of the interferometer. In fact, till now, we have just been
able to recover the correct shape of the interference pattern calculated
starting from the S-parameters of the full system. We remind that,
in the present analysis, the behavior of the optical part of the experimental
set-up has been considered Power6v1ideal. In the future, also an ad-hoc optical
bench for the far field illumination of the next generation of demonstrators
has to be designed, and a proper calibration strategy has to be foreseen.

\section*{Acknowledgements}
This work has been done in the frame of the BRAIN/MBI collaboration, the CNES\footnote{Centre des Etudes Spatiales du rayonnement}/CNRS\footnote{Centre National de la Recherche Scientifique} PhD funding for A. Ghribi and Ville de Paris Post-doc funding for A. Tartari. Special thanks to Alessandro Baù and Andrea Passerini for helping us in solving technical problems.
%\newpage{}
\bibliographystyle{IEEEtran}
\nocite{*}
\bibliography{dibo}

% Generated by IEEEtran.bst, version: 1.13 (2008/09/30)
\begin{thebibliography}{10}
\providecommand{\url}[1]{#1}
\csname url@samestyle\endcsname
\providecommand{\newblock}{\relax}
\providecommand{\bibinfo}[2]{#2}
\providecommand{\BIBentrySTDinterwordspacing}{\spaceskip=0pt\relax}
\providecommand{\BIBentryALTinterwordstretchfactor}{4}
\providecommand{\BIBentryALTinterwordspacing}{\spaceskip=\fontdimen2\font plus
\BIBentryALTinterwordstretchfactor\fontdimen3\font minus
  \fontdimen4\font\relax}
\providecommand{\BIBforeignlanguage}[2]{{%
\expandafter\ifx\csname l@#1\endcsname\relax
\typeout{** WARNING: IEEEtran.bst: No hyphenation pattern has been}%
\typeout{** loaded for the language `#1'. Using the pattern for}%
\typeout{** the default language instead.}%
\else
\language=\csname l@#1\endcsname
\fi
#2}}
\providecommand{\BIBdecl}{\relax}
\BIBdecl

\bibitem{key-1}
U.~Seljak and M.~Zaldarriaga, ``Signature of gravity waves in the polarization
  of the microwave background,'' \emph{Phys. Rev. Lett.}, vol.~78, no.~11, pp.
  2054--2057, Mar 1997.

\bibitem{key-2}
J.~Puget, ``Planck mission,'' \emph{35th COSPAR Scientific Assembly}, 2004.

\bibitem{key-3}
J.~Bock \emph{et~al.}, ``{Task Force on Cosmic Microwave Background
  Research},'' 2006.

\bibitem{key-4}
J.-C. {Hamilton}, R.~{Charlassier}, C.~{Cressiot}, J.~{Kaplan}, M.~{Piat}, and
  C.~{Rosset}, ``Sensitivity of a bolometric interferometer to the cosmic
  microwave backgroud power spectrum,'' \emph{A\&A}, 2008.

\bibitem{key-5}
M.~{White}, J.~E. {Carlstrom}, M.~{Dragovan}, and W.~L. {Holzapfel},
  ``{Interferometric Observation of Cosmic Microwave Background
  Anisotropies},'' \emph{The Astrophysical Journal}, vol. 514, pp. 12--24, Mar.
  1999.

\bibitem{key-6}
R.~{Charlassier for the BRAIN Collaboration}, ``The brain experiment, a
  bolometric interferometer dedicated to the cmb b-mode measurement,''
  \emph{Proceeding of the 43rd \char`\"{}Rencontres de Moriond\char`\"{} on
  Cosmology}, 2008.

\bibitem{key-7}
S.~Galli, ``Bolometric interferometry demonstrator to measure the cosmic
  microwave background polarization,'' AstroParticle \& Cosmology, Internship
  report, 2007.

\bibitem{key-8}
M.~J. Thompson, A.R. and J.~G. Swenson, \emph{Interferometry and Synthesis in
  Radio Astronomy}, Wiley-Interscience, Ed.\hskip 1em plus 0.5em minus
  0.4em\relax Wiley-Interscience, 2001.

\bibitem{key-9}
R.~{Charlassier}, J.-C. {Hamilton}, {\'E}.~{Br{\'e}elle}, A.~{Ghribi},
  Y.~{Giraud-H{\'e}raud}, J.~{Kaplan}, M.~{Piat}, and D.~{Pr{\^e}le}, ``{An
  efficient phase-shifting scheme for bolometric additive interferometry},''
  \emph{ArXiv e-prints}, Jun. 2008.

\bibitem{key-10}
E.~Bunn, ``Systematic errors in cosmic microwave background interferometry,''
  \emph{Physical Review D}, 2007.

\bibitem{key-11}
D.~M. Pozar, \emph{Microwave Engineering}, J.~W.~. Sons, Ed.\hskip 1em plus
  0.5em minus 0.4em\relax John Wiley \& Sons, 1998.

\bibitem{key-12}
A.~B.~B. Max~Born, Emil~Wolf, \emph{Principles Optics}, C.~U. Press, Ed.\hskip
  1em plus 0.5em minus 0.4em\relax Cambridge University Press, 1999.

\end{thebibliography}

\newpage{}

\section*{Appendix}

\subsubsection*{A.1 Relation between the field and the S parameters}

We consider here a two ports device. The source generator is connected
to port 1 and a matched load to port 2. We then have an incident wave
$V_{1}^{+}$ to the DUT (Device Under Test). The wave reflected from
the device back to port 1 is $V_{1}^{-}$. The signal traveling through
the DUT and toward port 2 is $V_{2}^{-}$. Any reflection from the
load is $V_{2}^{+}$. We can define S-parameters in terms of these
voltage waves \cite{key-11}:
\[
\begin{array}{cc}
S_{11}=\frac{V_{1}^{-}}{V_{1}^{+}} & S_{12}=\frac{V_{1}^{-}}{V_{2}^{+}}\\
S_{21}=\frac{V_{2}^{-}}{V_{1}^{+}} & S_{22}=\frac{V_{2}^{-}}{V_{2}^{+}}\end{array}\]
 where $S_{11}$ represents the input terminal reflection coefficient
$\Gamma_{1}$, $S_{21}$ forward gain or loss, $S_{12}$ Reverse gain
or loss and $S_{22}$ Output terminal reflection coefficient $\Gamma_{2}$.
We should underline that these parameters are normalized. Transient
field, is, by opposition to the tangent field, the one propagating
inside the waveguide. The tangent field is set to zero by the boundary
conditions of the Maxwell equation to avoid surface current in the
analysis.

We can express the transient E field in term of voltages:\[
E_{1}=\frac{e(x,y)}{c}\left(V_{1}^{+}e^{-i\beta_{1}}+V_{1}^{-}e^{i\beta_{1}}\right)\]
 \[
E_{2}=\frac{e(x,y)}{c}\left(V_{2}^{+}e^{-i\beta_{2}}+V_{2}^{-}e^{i\beta_{2}}\right)\]
 where $e(x,y)$ is the variation of the amplitude of the field propagating
inside a waveguide section and $c$ is a geometrical parameter. These
variables are set to constants because of the common waveguide shape
along the chain (cascade of components). $E_{1}$ and $E_{2}$ are the fields in the input
and in the output of the device and $\beta_{1}$ and $\beta_{2}$
are their respective phases. In order to express E2 with E1, we need
to consider the same base (the same propagation direction of the wave).
Let's set the propagation from port 1 to 2 as positive and from 2 to
1 as negative. This gives\[
E_{2}=\frac{e(x,y)}{c}\left(V_{2}^{+}e^{i\beta_{2}'}+V_{2}^{-}e^{-i\beta_{2}'}\right)\]
 with $\beta_{2}'=-\beta_{2}$

Finally, expressing $E_{2}$ as a function of $E_{1}$, we obtain\[
E_{2}=\left(\frac{S_{21}e^{-i\beta_{2}}+\left(\frac{S_{11}}{S_{12}}\right)e^{i\beta_{2}}}{e^{-i\beta_{1}}+S_{11}e^{i\beta_{1}}}\right)\times E_{1}\]

This relation is useful to make the link between the VNA measured
S parameters and the propagating field for a two ports device. Also,
it can be used to translate a 4 ports or double polarization propagation
from an S matrix formalism into a Jones matrix formalism.\\

\subsubsection*{A.2 Relation between the S parameters and the
ABCD parameters}

The ABCD formalism allows to propagate the signal over a chain of
two ports devices. The relation between the S parameters and ABCD
parameter is given in \cite{key-11}.

\end{document}